\documentstyle[prl,aps,twocolumn]{revtex}

\newcommand{\beq}{\begin{equation}}
\newcommand{\eeq}{\end{equation}}
\newcommand{\bea}{\begin{eqnarray}}
\newcommand{\eea}{\end{eqnarray}}

\begin{document}
\draft
%\tighten

\title{The Single-Electron-Box and the Helicity Modulus of an
inverse square XY-Model}

\author{W. Hofstetter and W. Zwerger}
\address{Sektion Physik,Ludwig-Maximilians-Univ.\ M\"{u}nchen,
Theresienstr.\ 37, D-80333 M\"{u}nchen, Germany}

\date{October 4, 1996}

\maketitle

\begin{abstract}
We calculate the average number of electrons on a metallic single-electron-box
as a function of the gate voltage for arbitrary values of the tunneling
conductance. In the vicinity of the plateaus the problem is equivalent 
to calculating the helicity modulus of a classical inverse square XY-model
in one dimension. By a combination of perturbation theory, a two-loop
renormalization group calculation and a Monte-Carlo simulation in the 
intermediate regime we provide a complete description of the smearing
of the Coulomb staircase at zero temperature with increasing conductance.
\end{abstract}

\narrowtext

In recent years the field of single electronics, which is based on the 
suppression of tunneling in small metallic islands or quantum dots 
by the Coulomb blockade, has gained enormous interest
\cite{Averin 91,Devoret 92}. The most 
elementary device where these effects can be studied is the so called
single-electron-box, first realized by Lafarge et al.\cite{Lafarge 91}. 
It consists of a small metallic island which is connected to an outside 
lead by a tunnel junction and is coupled capacitively to a gate voltage V. 
By elementary electrostatics the classical Coulomb energy
for a given integer number $n$ of additional electrons on the box is
\cite{Lafarge 91}
\beq
E_n = \frac{U}{2} (n-n_x)^2
\eeq
up to an irrelevant constant. Here $U = \frac{e^2}{C+C_g}$ is an effective 
single electron charging energy and $n_x = C_g V/e$ the continuous
polarization charge induced by the gate. Obviously, on varying $n_x$, the
actual integer value of $n$ is the one minimizing $E_n$. As a result, $n(n_x)$
is a staircase function with unit jumps at $n_x = \frac{1}{2}$ (mod 1).
The basic requirements for such a device to work are twofold:
First, it is obvious that the temperature $T$ (we set $k_B = 1 $) has to be
much smaller than the relevant charging energy. This  point is easily
taken into account by considering a thermal distribution of the energies
$E_n$ \cite{Lafarge 91}. The thermal average $\left< n \right>(n_x) $ 
will then approach
the simple straight line at $T \gg U$ (note that $n$ is measured over a 
sufficiently long time interval, giving a continuous 
$\left< n \right> $ even though the number of additional electrons
in the box is an integer at any given instant of time). 
Secondly, however, there is an intrinsic broadening of the staircase
even at $T=0$ since the tunneling probability through the junction 
is necessarily finite. As a consequence the number of electrons
in the box is not strictly conserved and the variable $n$ exhibits 
quantum fluctuations, which are neglected in a simple electrostatic 
description. A quantitative measure of these fluctuations is provided
by the average tunneling probability at the Fermi energy $\varepsilon_F$
which determines the dimensionless tunneling conductance\cite{Schoen 90}
\beq
g = \pi^2 |t|^2_{\varepsilon_F} \rho_{box} \rho_{lead} =: 
\frac{h}{4 e^2 R_t}
\eeq
Here $t$ is the transfer matrix element (see (\ref{hamil})) and $\rho$
are the densities of states at $\varepsilon_F$. Obviously the perfect
staircase at $T=0$ is only realized in the limit $g \rightarrow 0$,
while we expect that  $ \left<n\right> (n_x) \rightarrow  n_x$ even at 
zero temperature if $g \gg 1$. In the following we will give a quantitative
description of the behaviour of $\langle n \rangle(n_x)$ near the plateaus
at $n_x = 0$ (mod 1) for arbitrary values of g. It is shown that the slope
$\chi = \frac{\partial \left<n\right>}{\partial n_x}$ at $n_x = 0$ is
directly related to the helicity modulus of a classical 1-d inverse square
XY-model. Combining perturbation theory to order $g^2$, a two loop RNG
calculation for $g \gg 1$ and a MC simulation in the intermediate regime,
we are able to fully describe the crossover from perfect plateaus 
$\chi(g \rightarrow 0) \rightarrow 0 $ to the complete absence of the Coulomb 
blockade $\chi(g \gg 1) \rightarrow 1$. In particular the RNG result gives
a definitive solution to the long-standing controversy over the 
preexponential factor at strong coupling.

We start from a tunneling Hamiltonian\cite{Schoen 90}
\beq
\hat{H} = \frac{U}{2} (\hat{n} - n_x)^2 + \sum_{a b} t_{a b} c^\dagger_{a}
c_{b} + h.c. +\hat{H_0} 
\label{hamil}
\eeq
describing the Coulomb energy of the box and the transfer of electrons
between states $b$ and $a$ from the box (index $b$) to the lead (index $a$) 
and vice versa. The contribution $\hat{H_0}$ is the Hamiltonian 
of noninteracting 
Fermions on both sides of the junction which act as reservoirs. The 
Coulomb interaction is only incorporated by the classical capacitive
energy, which is a good description for metallic systems
\cite{Averin 91,Devoret 92}. In the 
following we shall employ an effective model for the thermodynamics
of the box which is obtained by integrating out the fermionic degrees of
freedom. Using a second order cumulant expansion in the transfer term,
which is appropriate in the experimentally relevant limit of a large 
number of conductance channels, the reduced partition function 
$Z = \mbox{Tr} \frac{\exp (-\beta \hat{H})}{Z_0}$ can be written as a
path integral\cite{Schoen 90}
\beq
Z(n_x) = \int^{\pi}_{-\pi} d\theta \int^\theta_\theta {\cal D} \theta
\exp \left\{ - S[\theta] +  i n_x \int_{-\frac{L}{2}}^\frac{L}{2} dx  
\frac{d\theta}{dx} \right\} 
\eeq
over a compact angular variable $\theta$ conjugate to the integer $n$.
Here $x$ is a dimensionless coordinate of a 1-d system with length 
$L = \beta U$ where $L \rightarrow \infty$ as the physical temperature
approaches zero. The action is given by $(-L/2 \le x \le L/2)$
\beq
S[\theta] = \frac{1}{2}\int dx \left(
\frac{d\theta}{dx}\right)^2 + \frac{2 g}{\pi^2} \int\!\int dx dx' 
\frac{\sin^2(\frac{\theta(x) - \theta(x')}{2})}{(x-x')^2} 
\label{action}
\eeq
The long range part of the interaction has already been written in the 
form appropriate for $L \rightarrow \infty $. Introducing a two component
unit spin ${\bf S}(x) = \left(\cos(\theta(x)) , \sin(\theta(x))  
\right) $ at any point of the line,
the action $S[\theta]$ is just the classical energy of an XY-model with 
an inverse square interaction proportional to the conductance $g$. 
The first term, which arises from the classical charging energy 
$\frac{C}{2} V^2 $ is then identical with the spin wave approximation
to a short range interaction. Finally, the external charge $n_x$ acts 
like a purely imaginary external torque on the XY-model. Since
$ \int d\theta = 2\pi m $ 
determines an integer winding number $m\in Z$ which is a topological 
invariant for each configuration $\theta(x)$, the external charge appears
as a pure boundary term.
Defining the free energy per length by 
$f(n_x) = -\ln Z(n_x)/L$ 
the average number of electrons in the box can be expressed as 
\beq
\langle \hat{n} \rangle = n_x - \frac{\partial f(n_x)}{\partial n_x}
\label{average}
\eeq
The fact that $n_x$ only arises as a phase factor $e^{2\pi i m n_x}$
shows that $Z(n_x +1) = Z(n_x)$ quite generally. Thus, all quantities
are periodic in $n_x$ with period one which allows to restrict the 
discussion to the intervall $-\frac{1}{2} < n_x \le \frac{1}{2} $.
In the following we confine ourselves to the zero temperature (i.e.
thermodynamic) limit $L \rightarrow \infty$ and to the vicinity of
$n_x = 0$. Taking $i n_x = m_x$ to be a real torque for the moment,
a finite value of $m_x$ will induce a nonzero average gradient of the phase
\beq
\lim_{L \rightarrow \infty} \frac{1}{L} \langle \int dx \frac{d\theta}{dx}
\rangle_{m_x} =  \frac{m_x}{\gamma}
\eeq
which is linear in $m_x$ in the limit $m_x \rightarrow 0$. The 
associated torsional rigidity $\gamma$ is then precisely identical
with the helicity modulus as defined by Fisher et al.\cite{Fisher 73}.
It may be obtained from $f(n_x)$ via
\beq
\frac{1}{\gamma} = \frac{\partial^2 f(n_x)}{{\partial n_x}^2} \Big|_{n_x = 0}
\eeq
which is a measure for the sensitivity to a change in the boundary conditions.
Using (\ref{average}) the slope of the Coulomb staircase near $n_x = 0$
is related to the helicity modulus by 
\beq
\chi = \frac{\partial \langle \hat{n} \rangle}{\partial n_x} \Big|_{n_x = 0}
= 1 - \gamma^{-1} 
\eeq
In the trivial limiting case $g = 0$ this describes
the expected result $\chi^{(0)} = 0$, i.e. perfect plateaus. 
Indeed if $g = 0$, the helicity modulus is equal to one, 
being just the coefficient in front of the $\frac{1}{2}\left(\frac{d\theta}{dx}\right)^2$ term\cite{Fisher 73}.
In order to describe the behaviour at finite $g$ we apply 
three different methods.

Perturbation Theory:
While the nonlinearity of the $\sin^2$ in the action (\ref{action}) makes an
exact evaluation of the path integral impossible, we may
expand the long range contribution down to second order in $g$.
Evaluating the resulting averages $\langle \exp(\pm i \theta ) \rangle $
with the remaining Gaussian action, the free energy can be calculated 
up to order $g^2$. After a straightforward but tedious calculation we 
obtain
\beq  \label{pert}
\chi(g) = c_1 g + c_2 g^2 + \ldots
\eeq
with $c_1 = \frac{4}{\pi^2}$ and $c_2 \approx -0.052 $. For the coefficient
$c_2$ we have evaluated a remaining definite integral numerically. The result
(\ref{pert}) is identical with a previous calculation by 
Grabert\cite{Grabert 94}, who has used direct 
perturbation theory to fourth order in $t$ in the original Fermionic 
Hamiltonian (\ref{hamil}). The agreement to order $g^2$ which we have verified
to 8 digits, confirms that the inverse square XY-model employed here is
a correct representation for the reduced thermodynamics of the original
model.

Renormalization Group:
To obtain the exact behaviour of the helicity modulus at large
values of $g$, we use the RNG. 
Indeed the limit $g \gg 1$ has previously been treated by approximate 
instanton calculations\cite{Panyukov 91,Wang 96}. However,
they give different results for the pre-exponential factor of the effective 
charging energy which is essentially the inverse of the helicity 
modulus. Here we will show that a definitive solution of this problem
may be obtained by a two loop RNG, which uniquely determines both the
exponent and the $g$-dependence of the prefactor of the correlation 
length $\xi$ in the limit $g \gg 1$. 
Indeed the helicity modulus is directly
proportional to the correlation length. To see this we define
\beq
\rho_m = 2\pi\:Z(n_x = 0)^{-1} \int_0^{2\pi m} {\cal D}\theta \exp \left\{
- S[\theta] \right\}
\label{rhodef}
\eeq
as the probability for a given winding number $m$. It is then straightforward
to show that the inverse helicity modulus 
\beq
\gamma^{-1} = \lim_{L \to \infty} \frac{\langle (2\pi m)^2 \rangle}{L}
\label{windav}
\eeq
measures the normalized variance of the winding number with respect to the 
probability distribution (\ref{rhodef}).
At \mbox{$g\gg 1$} a given value of $\langle m^2 \rangle$ requires a system
size which is larger by a factor of $\xi(g)$ than that at $g$ of order one
where $\xi(1) \sim O(1)$. Therefore, by applying (\ref{windav}) we have
$\gamma(g) \sim \xi(g) $. In order to determine $\xi(g)$, we use the fact
that $d=1$ is the lower critical dimension of the inverse square 
XY-model\cite{Kosterlitz 76,Brezin 76}. Since the kinetic energy term in 
(\ref{action}) is irrelevant
at \mbox{$g \gg 1 $} and $T_{eff} = \pi^2/g$ is the effective 
temperature
of our classical XY-model, we may perform a $d=1+\epsilon$ - expansion
around an ordered state at $\epsilon > 0$, which is effectively a low
temperature expansion. It is convenient to generalize the XY-spin 
${\bf S}(x)$ to a $O(n)$ spin $\mathbf{S}$ parametrized 
by\cite{Amit 84}
\beq
{\bf S}(x) = \left( {\bf \Pi}(x), \sqrt{1 - {\bf \Pi^2(x)}}\right)
\eeq
Here the $\Pi_i(x), i=1,...n-1$ are Goldstone modes whose expectation
scales like $\langle {\bf\Pi}^2 \rangle \sim T_{eff}$. We thus expand
the long range part of the action in powers of ${\bf\Pi}$. In Fourier
space and with $H$ as an external magnetic field which serves to regularize
the infrared divergencies, the action takes the form $(d=1)$
\beq
S =  \int dq \frac{|q|+H}{2 T_{eff}} \left[ {\bf\Pi}_q {\bf\Pi}_{-q}
+ \frac{1}{4} {\bf\Pi}^2_q {\bf\Pi}^2_{-q} + \frac{1}{8} {\bf\Pi}^2_q
({\bf\Pi}^2)^2_{-q} \right]
\eeq
up to terms of order $\Pi^8$. Following the method of 	Amit \cite{Amit 84} 
we calculate the $\beta$-function by field theoretic
renormalization using dimensional regularization. The two point function 
at zero external momentum is given by ($\epsilon \to 0$)
\bea \label{zweipunkt}
\Gamma^{(2)}(0,H) &=& \frac{H}{T_{eff}} - \frac{n-1}{2\epsilon} 
H^{1+\epsilon}  \\
&&+ T_{eff} H^{1+2\epsilon} \left[ \frac{(n-1)(n-2)}{4\epsilon} + 
\frac{3(n-1)^2}{8\epsilon^2} \right]  \nonumber  
\eea
in two loop order.
It can be made finite by introducing renormalized parameters
\cite{Brezin 76}
\beq
t = \kappa^\epsilon Z^{-1} T_{eff}\ \ , \ \ h = Z^{-\frac{1}{2}} H
\eeq
and fields
\beq
{\bf\Pi}_R = Z^{-\frac{1}{2}} {\bf\Pi}\ \  , \ \  \Gamma^{(2)}_R = 
Z \Gamma^{(2)} \ \ .
\eeq
The renormalization constant $Z$ turns out to be
\beq
Z = 1 + \frac{n-1}{\epsilon} t + \left[ \frac{(n-1)^2}{\epsilon^2} +
\frac{n-1}{2 \epsilon} \right] t^2 + O(t^3) 
\eeq
The resulting $\beta$-function for the renormalized temperature is then
given by 
\beq
\beta(t) = \epsilon t -(n-1)t^2 -(n-1)t^3 +O(t^4)
\eeq
This implies that under a reduction $\Lambda \to \Lambda \exp(-l)$ of the
cutoff the parameter $g^{-1} \sim T_{eff}$ at $\epsilon = 0$ scales like
\beq
\frac{d g^{-1}}{dl} = \frac{1}{2 g^2} + \frac{1}{4 g^3} + O(g^{-4})
\label{gflow}
\eeq
By integrating this differential equation, we find that the associated
correlation length diverges like
\beq    \label{correlation}
\xi(g \gg 1) = c(g) g^{-1} \exp(2 g)
\eeq
with a function $c(g)$ which is finite as $g \to \infty$. The exponential
behaviour is typical for a system at its lower critical dimensionality,
and is also obtained in an instanton approach\cite{Schoen 90,Kosterlitz 76}. 
However, the prefactor proportional
to $g^{-1}$ which is fixed by the coefficient of the two loop contribution
in (\ref{gflow}) and which implies that 
\beq   \label{asymptchi}
\chi(g \gg 1) = 1 - \bar{c} g \exp(-2 g)
\eeq
is quite different from the $g^2$\cite{Panyukov 91} or $g^3$\cite{Wang 96}
prediction of the instanton calculation. In fact, a very similar situation 
arises in the closely related $O(n)$ nonlinear $\sigma$-model in two
dimensions. Due to the scale invariance of the action there are instantons of
arbitrary size and the calculation cannot be controlled in the 
thermodynamic limit. It is only the two loop RNG which allows to
determine the correct prefactor of $\xi$\cite{Zinn 96}, although --
in contrast to the instanton results -- it does not fix the numerical
constant $\bar{c}$. 
It should also be pointed out that the finite correlation length
(\ref{correlation}) does not imply an exponential decay of 
$\langle {\bf S}(x) {\bf S}(0) \rangle $.
Indeed it can be shown \cite{Zwerger 91} that this correlation function
asymptotically decays like $1/x^2$ for all $0<g<\pi^2/2 $ and --
possibly -- even more slowly for larger values of $g$.  
For our discussion of the helicity modulus, however, the detailed 
behaviour of $\langle {\bf S}(x) {\bf S}(0) \rangle $ is irrelevant.

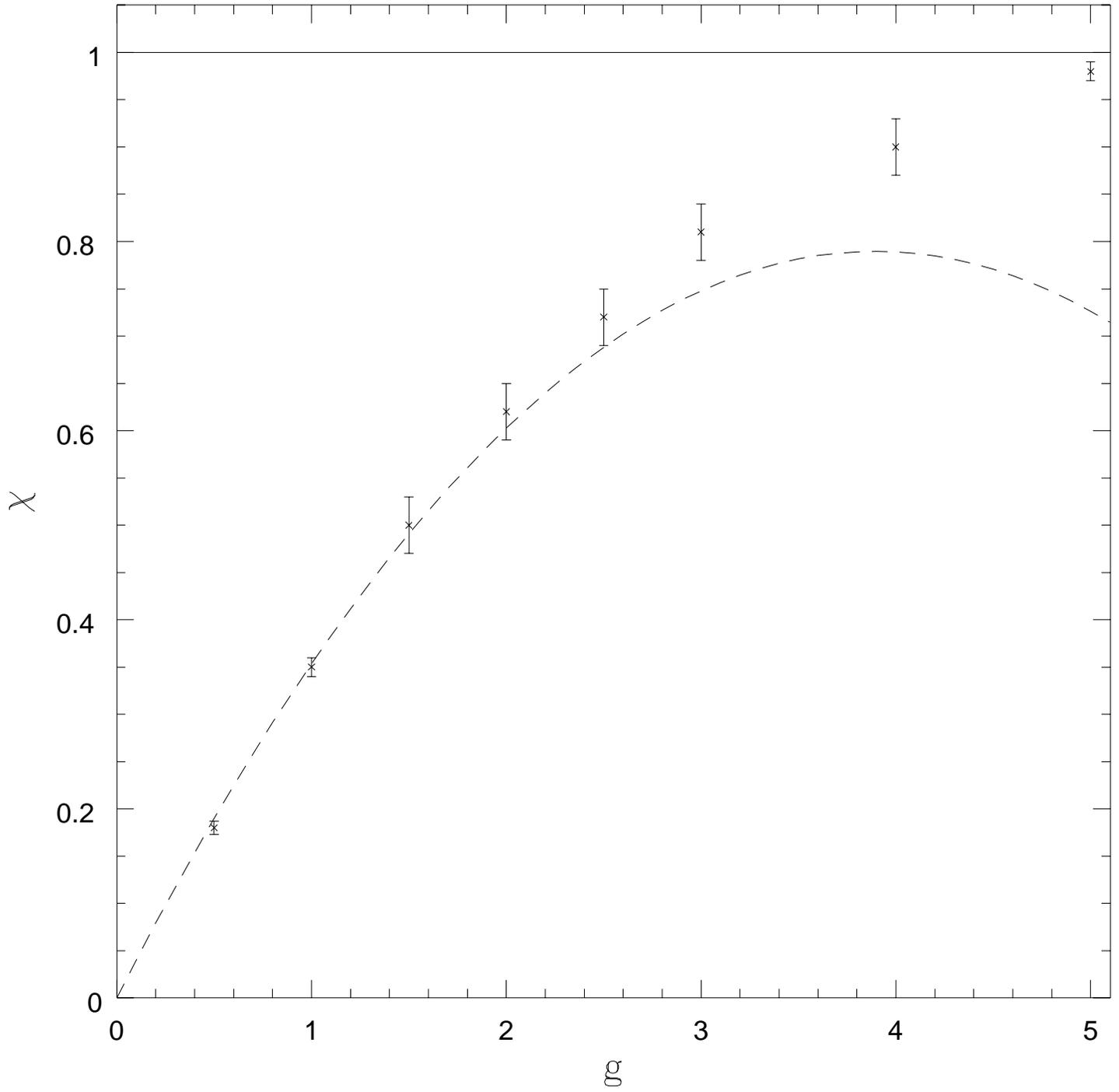
\begin{figure}
\caption{\label{fig:values} MC results for the zero temperature 
slope $\chi$ of the Coulomb
staircase at $n_x = 0$ as a function of the dimensionless conductance $g$.
The dashed line is the perturbative result (\ref{pert}).}
\end{figure}

Monte Carlo simulation:
In order to bridge the gap between the perturbative result (\ref{pert})
valid for small $g$ and the asymptotic behaviour (\ref{asymptchi}) we
have performed a Monte Carlo simulation of the inverse square XY model
(\ref{action})
including the short range interaction term for values of $g$ between 
0.1 and 5. 
Following previous work\cite{Scalia 91,Falci 95} we have sampled the 
winding number probabilities
$\rho_m$ defined in (\ref{rhodef})  
using the standard Metropolis
algorithm with periodic boundary conditions on a discrete chain with 
up to 2000 spins.
We have checked carefully that further increase in the system length
does not change our results for the helicity modulus.
To estimate the statistical error
we have performed about 40 runs for each choice of the paramters $g$ and $L$.
The calculations were done with HP 700 workstations and a CRAY T90 and took
about one hour of CPU time per run.
The numerical data are shown in Fig. \ref{fig:values}. Evidently the slope of 
$\langle \hat{n}\rangle$ versus $n_x$ follows 
the perturbative result (\ref{pert})
closely up to values around $g \approx 1$ and finally approaches 
$\chi = 1$
exponentially fast as predicted by (\ref{asymptchi}). 
Assuming that the asymptotic behaviour (\ref{asymptchi}) is already 
valid for $g > 3$, the MC results allow to determine the constant
in (\ref{asymptchi}), giving $\bar{c} = 80$ from a two point fit.
One should note that a prefactor of this magnitude has also been obtained
for the low-temperature-behaviour of the correlation length
of the nonlinear $\sigma$-model \cite{Shenker 80}.
It is difficult to estimate, however, whether the asymptotic 
behaviour has already been reached at these values of $g$. Unfortunately
the exponential increase of the correlation length does not allow 
us at present to verify numerically our analytical result 
(\ref{asymptchi}) for the $g$-dependence of the prefactor\cite{Egger 96}.
Indeed the situation is again analogous to the much studied $O(n)$
nonlinear $\sigma$-model in $d=2$. While the purely numerical 
constant equivalent to $c(\infty)$ in (\ref{correlation}) was 
determined approximately by Shenker and Tobochnik\cite{Shenker 80}
via a Monte-Carlo RG method, the exact coupling constant dependence
of the prefactor which follows from the two loop RNG\cite{Brezin2 76} has 
only been verified in recent years by extremely extensive numerical 
computations\cite{Drouffe 90}. Since the range of physical interest 
in the single electron box problem is restricted to $g$ values smaller
than about five (i.e. $R_t \ge 1.3\mathrm{k}\Omega$, see Fig. 1) it is evident 
that our present numerical results fully cover the experimentally 
accessible regime.

In conclusion we have calculated the zero temperature smearing of the 
Coulomb staircase in the single electron box for arbitrary values of
the tunnel conductance $g$. In contrast to previous work on this
problem, which has concentrated on the behaviour near $n_x = 1/2$
and $g \ll 1$ \cite{Matveev 91} we have discussed the slope at the center 
of the plateaus. 
It has been shown that this is determined
by the helicity modulus of an inverse square XY-model. 
Quantitative MC simulations in the physically 
relevant regime $g = 0.1 \ldots 5$ yield the expected crossover between 
perturbation theory and the asymptotic behaviour. 
Moreover, the two loop RNG calculation uniquely determines the analytical 
behaviour at large conductance and shows that previous instanton 
calculations are problematic. In the existing experiments 
\cite{Lafarge 91} the conductance had a fixed value of order $g_{exp}
\approx 0.02 $, i.e.\ well in the perturbative regime. In fact the 
first oder correction in (\ref{pert}) was used to calibrate 
the slope at $n_x = 0$ to zero. To verify our results one therefore
needs measurements with different and considerably larger 
values of $g$ at temperatures where the thermal broadening 
is negligible.

\end{document}